\documentclass[10pt]{iopart}
\usepackage[final]{graphics}
\usepackage{amssymb}
\usepackage{amsfonts}
\usepackage{epsfig}
\usepackage{graphicx}

\begin{document}

\title{Universality of the contact process with random dilution}

\author{Marcelo M. de Oliveira} \ead{mancebo@fisica.ufmg.br}
\author{Silvio C. Ferreira} \ead{silviojr@ufv.br}
\address{Departamento de F\'{\i}sica, Universidade Federal
Vi\c{c}osa, 36571-000, Vi\c{c}osa, MG, Brazil}

\date{\today}

\begin{abstract}
We present quasi-stationary simulations of the two-dimensional
contact process with quenched disorder included through the random
dilution  of a fraction of the lattice sites (these sites are not susceptible to infection).  Our results strongly indicate
that the static exponents are independent of the immunization fraction.  In addition, the critical moment ratios $m=\langle\rho^2\rangle/\langle\rho\rangle^2$  deviate from the universal ratio $m=1.328$, observed for the non-dilluted system, to smaller values due to rare favorable regions which dominate the statistics.

\end{abstract}


\submitto{\it J. Stat. Mech.}

\section{Introduction}

Absorbing-state phase transitions, i.e, transitions from a
fluctuating phase to an absorbing (trapped) state, are related to
several nonequilibrium critical phenomena
\cite{marro,hinrichsen,odor04,lubeck} such as chemical catalysis
\cite{ziff86}, interface growth \cite{tang}, epidemic spreading
\cite{bartlett}, etc. The study of such transitions in spatially
extended systems has been experimenting an ongoing interest,
strengthened by recent experimental confirmations of absorbing-state
phase transitions in a liquid crystal system \cite{take07}, and in a
sheared colloidal suspension \cite{pine}. Notwithstanding a complete
classification of their critical behavior is still missing, it has
been conjectured \cite{jans81,gras82} that models with a positive
one-component order parameter, short-range interactions and deprived
of additional symmetries or quenched disorder belong generally to
the universality class of directed percolation (DP), which is
considered the most robust universality class of the absorbing-state
phase transitions.

The contact process (CP) \cite{harris-CP} is one of the simplest and
most studied models of the DP universality class. Of particular
interest is how spatially quenched disorder affects its critical
behavior \cite{cafiero,webman}. Quenched disorder, in the form of
impurities and defects, plays an important role in real systems, and
may be responsible for the rarity of experimental realizations, in
spite of the ubiquity of the DP class \cite{hinri00b}.

The so-called Harris' criterion \cite{harris74} states that quenched
disorder is a relevant perturbation, from the field-theoretical
point of view, if
\begin{equation}
d\nu_\perp<2,
\end{equation}
where $d$ is the dimensionality and $\nu_\perp$ is the correlation
length exponent of the pure model (In DP this inequality is
satisfied in all dimensions $d<4$, since $\nu_\perp =$ 1.096854(4),
0.734(4) and 0.581(5), for $d=1$, $2$ and $3$, respectively
\cite{jensen99,voigt97,jensen92}). The first numerical studies of
the CP with quenched disorder, introduced by the means
of a random deletion of sites \cite{noest,adr-dic} or bonds \cite{dahmen}
(dilution) or by random random spatial variation of the
control parameter  \cite{durrett}, confirmed that the disordered
system does not belong to the DP class \cite{noest}, and also
revealed that the critical spreading is logarithmic, not a power law
\cite{adr-dic}. In the subcritical regime, a Griffits-phase, with
critical dynamics dominated by nonuniversal power laws was also
reported \cite{adr-dic,cafiero}. However, the contact process in a
Voronoi-Delaunay lattice, which has an intrinsic quenched disorder in the
distribution connectivity, belongs to the DP class \cite{Oliveira2008}.

Hooyberghs et al. \cite{hooy,hooy2}, employed a
strong-disorder renormalization group approach to conclude that the
unusual critical behavior of the disordered system can be related to
the random transverse-field Ising model, for sufficiently strong
disorder. At such infinite-randomness fixed point, the scaling is
activated, i.e, the temporal and spatial correlation lengths
($\xi_\parallel$ and $\xi_\perp$, respectively) are related by
\begin{equation}
\ln{\xi_\parallel}\sim\xi_\perp^\psi
\end{equation}
where $\psi$ is a universal exponent. For weak disorder they found
nonuniversal critical exponents depending on the disorder strenght.

More recently, Votja and Lee \cite{vojta06} used this activated
dynamic scaling to show that the interplay between geometric
criticality and dynamic fluctuations leads to a novel universality
class, with the exponent $\psi$ equals to the fractal dimension of
the critical percolation cluster of the diluted lattice. Previous
numerical studies of the one-dimensional contact process with
quenched spatial disorder, performed by Votja and Dickinson
\cite{vojta05} also supported the activated exponential dynamical
scaling at the critical point. Moreover, they found evidences that
this critical behavior turns out to be universal, even for weak
disorder. Novel strong disorder renormalization group calculations
in a very recent paper by Hoyos \cite{hoyos} predict that the system
is driven to the infinite-randomness fixed point, independently of
the disorder strength.

Thus, despite of a deeper understanding of the effects of quenched
disorder in the critical contact process achieved in the last
decade, a certain controversy remains: Do the static critical
exponents change continuously with the degree of disorder
\cite{adr-dic,cafiero,hooy}, or do they change abruptly to the
values in the strong disorder limit corresponding to the
universality class of the random transverse Ising model
\cite{vojta05,vojta06}? Obtaining the static exponents is a hard
numerical task because at criticality the infinite disorder fixed
point is characterized by a ultra-slow dynamics, $\rho(t)\sim
\ln(t)^{-\beta/(\nu_\perp\psi)}$, leading to a unusual long
relaxation towards the quasi-stationary (QS) values. Thus, in a
tentative to shed some light into this issue, we
present results of extensive large-scale QS simulations \cite{qssim}
of the two-dimensional diluted contact process.

The balance of this paper is organized as follows: In the next
section we review the definition of the contact process and describe
the simulation method. In Sec. III we present our
results and discussion, and Sec. IV is devoted to draw some conclusions.

\section{Model and methods}

The contact process is a stochastic interacting particle process in
which the particles lie on the sites of a $d$-dimensional hypercubic
lattice. Each site can be vacant or occupied. An empty site becomes
occupied at a rate $\lambda n/d$, where $n$ is the number of its
occupied nearest-neighbors, while occupied sites become vacant at
unitary rate \cite{harris-CP,marro}. For a certain critical value
$\lambda=\lambda_c(d)$, the model exhibits a continuous phase
transition from an active state (with a positive density of sites
occupied) to an absorbing configuration with all sites vacant, since
none particle can be created from the vacuum.

In the simulation we employ the usual scheme \cite{marro}: with
probability $p=\lambda/(1+\lambda)$, one nearest neighbor $j$ of the
selected site $i$ is chosen at random and occupied if the site $j$
is vacant. With complementary probability $q=1/(1+\lambda)$ the
particle at site $i$ is annihilated. At each step, the time is
increased by $\Delta t =1/N_{occ}$, where $N_{occ}$ is the total
number of occupied sites. Moreover, occupied sites are sequentially
selected at random from a list constantly updated, in order to
improve efficiency.

In the original diluted contact process (DCP) \cite{noest,adr-dic},
a quenched disorder is introduced by labeling each site as diluted
or nondiluted with probabilities $x$ and $1-x$, respectively. In the
present work, we mark as diluted exactly a fraction $\Gamma$ of the
sites, in order to avoid undesirable extra fluctuations. Thus, DCP
is the CP model restricted to the nondiluted sites since those
diluted are never occupied. Notice that, for large systems, the
dilution process used in this work is equivalent to those used in
reference \cite{adr-dic}.

Stationary analysis nearby the critical point of systems with
transitions to absorbing configurations are hard to be done due to
very strong finite size effects. Indeed, the unique real stationary
state of finite systems is the absorbing configuration. A common
alternative to avoid this difficulty is to restrict the averages to
the surviving samples and proceed a finite size analysis. This
procedure \cite{marro}, involves careful scrutiny in the data
analysis which is not always free of ambiguities or
misinterpretations \cite{lubeck}. In order to circumvent such
difficulties, we employ a simulation method that yields
quasistationary (QS) properties directly, the QS simulation method
\cite{qssim}. The method is based in maintaining, and gradually
updating, a list of $M$ configurations visited during the evolution;
when a transition to the absorbing state is imminent the system is
instead placed in one of the saved configurations. Otherwise the
evolution is exactly that of a conventional simulation. By this
procedure one obtains an unbiased sampling of the quasistationary
distribution of the process.

The simulations were performed as follows. Firstly, the list of
configurations is incremented whenever the time increases by a unity
up to a list with $M=1000$ configurations is achieved. Secondly, a
configuration of the list is randomly chosen and replaced by the
current one with a given probability $p_{rep}$. We used a large
value of $p_{rep}=0.5$ for a initial relaxation period, more
precisely for $t<L^{1.5}$ where $L$ is the linear system size, in
order to speed up the erasing of the memory of the initial
condition. Also, $p_{rep}=0.005$ was adopted for the remaining of
the simulation.

\section{Results and Discussion}

We studied lattice sizes varying from $L=20$ to $L=640$, averaging
over 200 to 300 independent realizations of disorder, of duration up
to $t=2\times 10^8$. Averages were taken after a relaxation time
$t_r=10^8$. The larger sample sizes and longer run times apply to
the larger $L$ values.

The first step in analyzing our results is to determine, for each
dilution value studied, the critical creation rate
$\lambda(\Gamma)$. For this purpose we study the number of active
particles, $n(t)$ via spreading analysis. The critical value
$\lambda_c$ is then defined as the smallest $\lambda$ supporting
asymptotic growth (vide Fig.1). This criterion avoids
misinterpretations associated to the effects due to the Griffiths
phase, in which power laws in $n(t)$ are observed \cite{adr-dic}.

\begin{figure}[hbt]
\begin{center}
\includegraphics[width=8.0cm,height=!,clip=true]{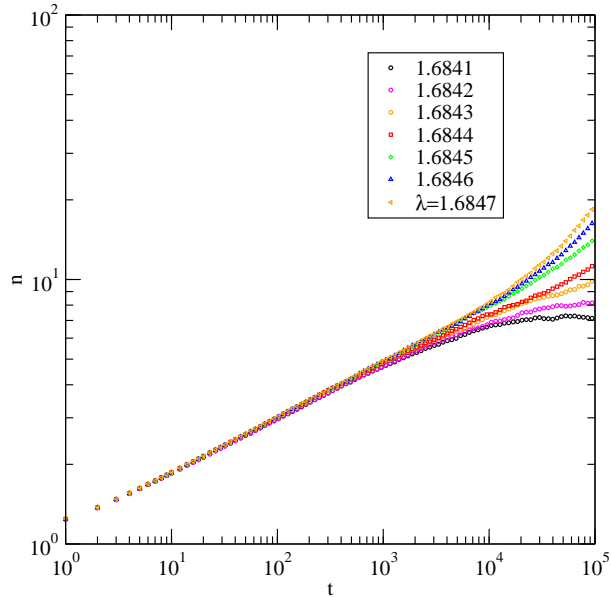}
\caption{\label{fig:esp} Spreading of activity from a single seed.
Dilution rate: $\Gamma=0.02$.}
\end{center}
\end{figure}

In Fig. 2 we show a log-log plot of the critical quasistationary
density of active sites $\rho$, as function of $L$ for dilutions
ranging from 0.02 to 0.30. At criticality, such quantity decays as a
power law, $ \rho \sim L^{-\beta/\nu_\perp}$. (This permits us to
check the critical values $\lambda_c$ obtained from the spreading
analysis). The values for the critical exponent $\beta/\nu_\perp$,
obtained from linear least-squares fits to the last four points of
the data are shown in Table I. Our results suggest that the critical
exponent ratio $\beta/\nu_\perp$ is {\it independent} of the amount
of dilution, with $\beta/\nu=0.95(2)$, at least for dilutions
$\Gamma\geq 0.05$.

\begin{figure}[hbt]
\begin{center}
\includegraphics[width=10.0cm,height=!,clip=true]{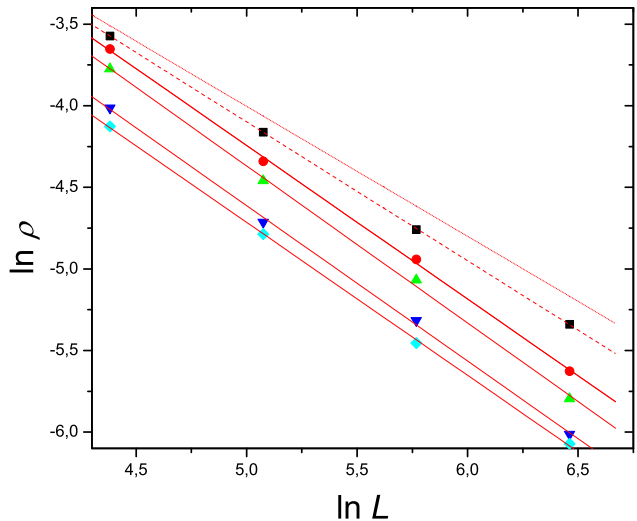}
\caption{Quasistationary densities of active sites versus system
size $L$. From top to bottom: nondiluted CP (dotted line),
$\Gamma=0.02$ (dashed), and $\Gamma=0.05, 0.10, 0.20$ and $0.30$.}
\end{center}
\end{figure}

\begin{table}[!hbt]
\caption{Exponent ratio $\beta/\nu_\perp$ for several dilution
values. $^{a)}$Present work using $L\le 640$. $^{b)}$Taken from
\cite{adr-dic} where $L\in[8,128]$ was used.}
\begin{center}
\begin{tabular}{cccc}
\hline\hline $\Gamma$  & $\lambda_c$ &~~$^{a)}\beta/\nu_\perp~~$  &
$~~^{b)}\beta/\nu_\perp~~$  \\ \hline
0         & 1.64874(1) & 0.797(2)             & 0.80  \\
0.02      & 1.6844(1) & 0.87(1)              & 0.83 \\
0.05      & 1.7410(1) & 0.94(2)              & 0.82  \\
0.10      & 1.84640(5) & 0.96(2)             & 0.85  \\
0.20      & 2.1075(2)  & 0.95(2)              & 0.86  \\
0.30      & 2.473(1)  &  0.95(1)                & 0.92  \\
\hline
\end{tabular}
\end{center}
\end{table}

Now we turn to the dynamical exponent $z=\nu_\parallel/\nu_\perp$.
In the nondiluted CP, the lifetime of the QS state (which we take as
the mean time between two attempts to absorbing in the QS
simulation), follows $\tau\sim L^z$. On the other hand, in the
activated dynamical scenario, such power-law scaling is replaced by
$\ln \tau\sim L^\psi$, with $\psi$ being an universal exponent. In
other words, the critical exponent $z$ is formally infinity in this
scenario. Early works \cite{adr-dic, hooy} found a nonuniversal
power-law behavior, with an exponent $z$ varying continuously in
direction to the strong disorder values. Otherwise, our results,
shown in Fig.3, reveal that at criticality the lifetime behavior is
not a power law, clearly diverging with an increasing slope for all
$\Gamma\geq 0.05$. This suggest that the activated scaling emerge
even for weak disorder, enforcing the universality hypothesis.
Furthermore, applying the activated scaling to the the data for the
last four points for highest dilution, $\Gamma=0.30$ furnishes the
value of $\psi=0.48(7)$, consistent with the values of the exponent
$\psi$ in the range $0.42<\psi<0.50$ found in the literature for the
random transverse Ising model \cite{psi1,psi2}. For weaker disorder,
our results do not permit to distinguish between a scenario with
continuously varying $\psi$ to a possible crossover to a universal
value at larger system sizes, as predicted by the activated
dynamics.

\begin{figure}[hbt]
\begin{center}
\includegraphics[width=12.0cm,height=!,clip=true]{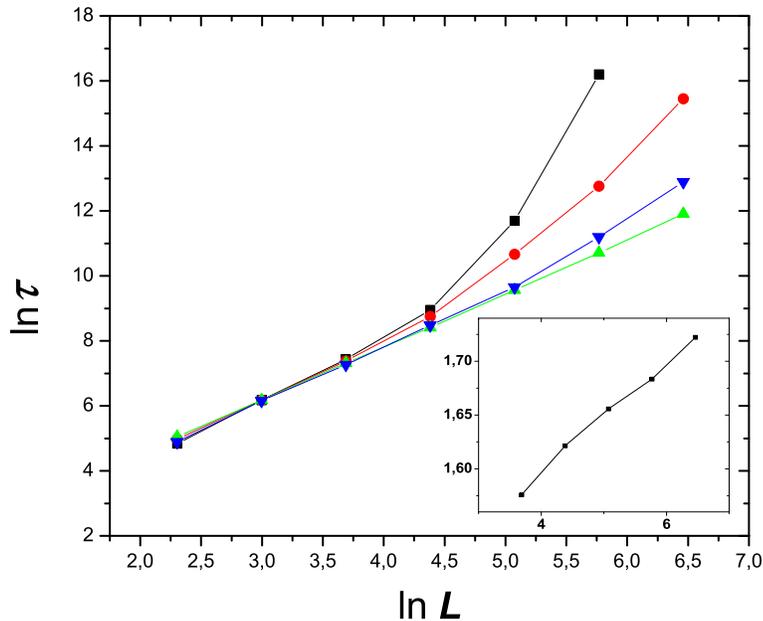}
\caption{ Quasistationary lifetime $\tau$ versus system size $L$.
Dilution rates: $\Gamma=0.05,0.10,0.20,0.30$, from bottom to top.
Inset: $d\Gamma/dL$ for $\Gamma=0.05$.}
\end{center}
\end{figure}

\begin{figure}[h]
\begin{center}
\includegraphics[width=12.0cm,height=!,clip=true]{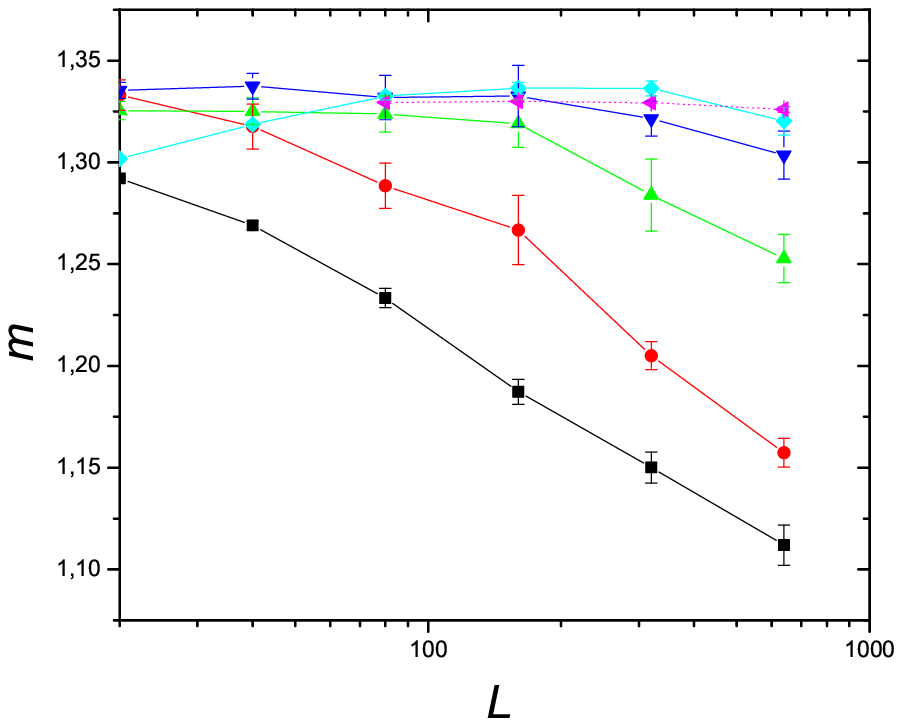}
\caption{Moment ratio $m$ for dilutions $\Gamma=0.30,0.20,0.10,0.05$
and $0.02$ from bottom to top. The dashed line represents the
nondiluted $\Gamma=0.00$ case.}
\end{center}
\end{figure}

Another consequence of the activated dynamics scenario is that the
distribution of the observables become broader, implying that the
averages are dominated by the rare events in which the process is
locally supercritical. Thus, some quantities such as moment ratios
of the order parameter (which converges to universal values in the
nondiluted CP \cite{dic-jaf}) exhibit non-self-averaging properties,
in the sense that they do not converge to limiting values even when
$L\to\infty$ \cite{adr-dic,wise,ahar}. This is exemplified in Fig. 4
where the moment ratio $m=\langle\rho^2\rangle/\langle\rho\rangle^2$
is plotted as a function of the system size $L$. The effects of the
rare `favorable' regions dominate the statistics, and the moment
ratio is drifted from the DP value of $m\sim 1.328$ to smaller
values (in the favorable regions the process is locally
supercritical, and $m\to 1$ in the limit $\lambda\to \infty$). We
observe that for high dilutions the effects of the rare regions
become observable even for modest system sizes, while for low
dilution the effect only appear at considerable larger system sizes.
Notice that for the dilution $\Gamma=0.02$ these effects are not
evident for the system sizes we used, what may justify the
difference in the exponent ratio $\beta/\nu_\perp$ for the smallest
value of $\Gamma$.

\section{Conclusions}

We performed extensive large-scale simulations of the
two-dimensional contact process with dilution. The dilution is known
to change the critical behavior of the contact process. Our results
indicate that the novel static exponents do not depend on the amount
of dilution, and we present numerical evidences that the apparent
nonuniversality observed in early works was due to finite-size
effects. Our findings are in agreement with recent simulational
results for the one-dimensional contact process with quenched
disorder \cite{vojta06}, and with strong disorder renormalization
group results \cite{hoyos}. On the other hand, our results cannot
exclude a nonuniversal variation of the exponent $\psi$ with the
disorder strength. Finally, the critical moment ratios
$m=\langle\rho^2\rangle/\langle\rho\rangle^2$ deviate from the the
universal ratio $m=1.328$  of the non-diluted system to smaller
values, due to rare favorable regions which dominate the statistics.

\bigskip

{\bf Acknowledgments}

We thank Ronald Dickman for helpful discussions
and for pointing us reference \cite{hoyos}. This work was partially
supported by CNPq and FAPEMIG Brazilian agencies.


\begin{thebibliography}{99}



\bibitem{marro}
        Marro J and Dickman R, 1999
        {\it Nonequilibrium Phase Transitions in Lattice Models}
        (Cambridge University Press: Cambridge).

\bibitem{hinrichsen}
     Hinrichsen H, 2000
     Adv. Phys. {\bf 49} 815.

\bibitem{odor04}
        \'Odor G, 2004
        Rev. Mod. Phys {\bf 76} 663.

\bibitem{lubeck}
        Lubeck S, 2004
J. Mod. Phys. B {\bf 18} 3977.

\bibitem{ziff86} Ziff R M , Gulari E, and Barshad Y, 1986 Phys. Rev. Lett. {\bf 56}
2553.

\bibitem{tang} Tang L H and Leschhorn H, 1992 Phys. Rev. A {\bf 45}
R8309.

\bibitem{bartlett}
        Bartlett M S, 1960
        {\it Stochastic Population Models in Ecology and Epidemiology}
        (Methuen: London).

\bibitem{take07}
     Takeuchi K A, Kuroda M, Chat\'e H, and Sano M, 2007
     Phys. Rev. Lett. {\bf 99} 234503.

\bibitem{pine}
Cort\'e L, Chaikin P M, Gollub J P, and Pine D J, 2008 Nature
Physics {\bf 4} 420.

\bibitem{jans81}
Janssen H K, 1981 Z. Phys. B {\bf 42} 151.

\bibitem{gras82}
Grassberger P, 1982 Z. Phys. B {\bf 47} 365.

\bibitem{harris-CP}
         Harris T E, 1974
        Ann. Probab. {\bf 2} 969.

\bibitem{cafiero}
Cafiero R, Gabrielli A, and Munoz M A, 1998 Phys. Rev. E {\bf 57}
5060.

\bibitem{webman}
Webman I, ben-Avraham D, Cohen A, and Havlin S, 1998 Phil. Mag. B
77, 1401.

\bibitem{hinri00b}
Hinrichsen H, 2000 Braz. J. Phys. {\bf 30} 69.

\bibitem{langhorne} Langhorne J, Ndungu FM, Sponaas A, and Marsh K, 2007 Nature Immun. \textbf{9} 725.

\bibitem{harris74}
        Harris A B, 1974
        J. Phys. C {\bf 7}, 1671.

\bibitem{jensen99} Jensen I, 1999 J. Phys A {\bf 32} 5233.

\bibitem{voigt97} Voigt C A and Ziff R M, 1997 Phys. Rev. E {\bf 56} R6241.

\bibitem{jensen92} Jensen I, 1992 Phys. Rev. A {\bf 45} R563.


\bibitem{noest}
Noest A J, 1986 Phys. Rev. Lett. {\bf 57} 90; \\
Noest A J, 1988 Phys. Rev. B {\bf 38} 2715.

\bibitem{adr-dic}
Moreira A G and Dickman R, 1996 Phys. Rev. E \textbf{54} R3090; \\
Dickman R and Moreira A G, 1998 Phys. Rev. E \textbf{57} 1263.

\bibitem{dahmen} Dahmen S R, Sittler L, and Hinrichsen H, 2007 J. Stat. Mech P01011.

\bibitem{durrett} Bramson M, Durrett R, and Schonmann R, 1991 Ann. Prob. {\bf 19},
960.

\bibitem{Oliveira2008} de Oliveira M M, Alves S G, Ferreira S C, Dickman R, 2008 Phys. Rev. E {\bf 78}, 031133.

\bibitem{hooy} Hooyberghs J, Igl\'oi F, and Vanderzande C, 2003 Phys. Rev. Lett.
{\bf 90} 100601.

\bibitem{hooy2} Hooyberghs J, Igl\'oi F, and
Vanderzande C, 2004 Phys. Rev. E {\bf 69} 66140.

\bibitem{vojta06}
Vojta T and Lee M Y, 2006 Phys. Rev. Lett. {\bf 96}, 035701.

\bibitem{vojta05} Vojta T and Dickison M, 2005 Phys. Rev. E \textbf{72}, 036126.

\bibitem{hoyos} Hoyos J A, 2008 Phys. Rev. E \textbf{78}, 032101.

\bibitem{qssim}
     de Oliveira M M and Dickman R 2005,
     Phys. Rev. E {\bf 71} 016129; \\
     Dickman R and de Oliveira M M, 2005
     Physica A {\bf 357} 134.

\bibitem{psi1}
 Motrunich O et al., 2000 Phys. Rev. B {\bf 61} 1160.

\bibitem{psi2}
Karevski D et al., 2001 Eur. J. Biochem. {\bf 20} 267.

\bibitem{dic-jaf}
     Dickman R and Kamphorst Leal da Silva J, 1998
     Phys. Rev. E {\bf 58} 4266.

\bibitem{wise} Wiseman S and Domany E, 1995 Phys. Rev. E {\bf 52} 3469.

\bibitem{ahar} Aharoni A and Harris A B, 1996 Phys. Rev. Lett. {\bf 77} 3700
.




\end{thebibliography}
\end{document}